\documentclass[]{article}

\usepackage[margin=0.7in]{geometry}
\usepackage[parfill]{parskip}
\usepackage[utf8]{inputenc}
\usepackage{gensymb}
\usepackage{textcomp}
\usepackage{graphicx}
\usepackage{hyperref}
\usepackage{placeins}
\usepackage{crop,inputenc,amsmath,amssymb,latexsym,
                float,epsfig,wrapfig,graphicx,stfloats,
                multicol,rotating,multirow,dcolumn,
                boxedminipage,makeidx,soul,url,xspace,times}

\begin{document}

\title{Simultaneous sampling of multiple transition channels using adaptive paths of collective variables}

\author{%
Alberto P\'erez de Alba Ort\'iz\,$^{1,2}$
\footnote{Email: a.perezdealbaortiz@uva.nl}
and Bernd Ensing\,$^{1,3}$%
}

\date{%
\small{$^{1}$Computational Chemistry, Van 't Hoff Institute for Molecular Sciences and Amsterdam Center for Multiscale Modeling, University of  Amsterdam, Science Park 904, 1098 XH Amsterdam, The Netherlands.\\
$^{2}$Computational Soft Matter, Van 't Hoff Institute for Molecular Sciences and Informatics Institute, University of  Amsterdam, Science Park 904, 1098 XH Amsterdam, The Netherlands..\\
$^{3}$AI4Science Laboratory, University of Amsterdam, Science Park 904, 1098 XH Amsterdam, The Netherlands.\\[2ex]
\today}
}


\maketitle

\begin{abstract}
We present a molecular simulation method to simultaneously find multiple transition pathways, and their associated free-energy profiles.
The scheme extends path-metadynamics (PMD) [Phys. Rev. Lett. 109, 020601 (2012)] and multiple-walker PMD [J. Chem. Phys. 149, 072320 (2018)] with multiple paths and repulsive walkers (multiPMD).
We illustrate multiPMD for two C7$_\text{eq}\rightarrow$C7$_\text{ax}$ paths in Ace-Ala-Nme and six PPII$\rightarrow$PPII paths in Ace-(Pro)$_4$-Nme.
\textcolor{black}{We also show a scheme to render an interpretable ``PathMap'', showing the free energy ridges between paths, as well as the branching and merging of the transition channels.}
MultiPMD is a flexible and promising method for systems with competing or controversial pathways,
\textcolor{black}{which appear in many biomolecular systems, including proteins and nucleic acids.}
\end{abstract}

\section{Introduction}

\textcolor{black}{In molecular simulations, enhanced sampling methods provide an effective means to deal with molecular transitions that are too slow or infrequent for direct simulation, such as chemical reactions, conformational rearrangements, and self-assembly processes.}
In particular, schemes that use a biasing force to escape the stable states and estimate the free energy---including umbrella sampling \cite{torrie_1977_us}, constrained molecular dynamics \cite{Carter1989:472,Otter1998:4139}, steered molecular dynamics \cite{grubmuller_1996_smd,jarzynski_1997_smd}, metadynamics \cite{laio_2002_mtd}, and many others \cite{Huber1994:695,Grubmuller1995:2893,Voter1997:3908,Darve2001:9169,Babin2008:134101,Darve2008_ABF,mohr2024enhanced}---depend crucially on an accurate description of the transition by one or more collective variables (CVs).
\textcolor{black}{However, the computational cost of sampling multidimensional free-energy landscapes scales exponentially with the number of CVs, and is therefore usually limited to a maximum of three.
To sample complex transitions that require more CVs to be described, path-CV approaches have been developed.}
A path-CV defines a mechanistic route from a reactant state to a product state as a parametrized curve in the space spanned by a set of standard CVs, e.g.\ bond lengths, angles, dihedral angles, or more elaborate  functions of the atomic coordinates. By optimizing this curve towards the minimum free-energy pathway (MFEP) or the mean transition pathway, the path-CV gives unique insight into the reaction coordinate of the slow process. Additionally, due to the subexponential scaling of the computational cost with the number of CVs in which the path-CV is expressed \cite{perezdealbaortiz_2018_pmtdadv}, more complex processes can be modeled than when the CVs are biased directly in an enhanced sampling simulation. The main disadvantage of the path-CV approach is however, that the reaction channel that it converges to depends on the starting location of an initial guess path, which poses a difficulty when the stable states are connected by more than one (possibly unknown) channel.
\textcolor{black}{In practice, multiple transition pathways are present in many biomolecular systems, e.g.\  protein folding (see \cite{eaton2017multiprotein} and references therein), small molecule unbinding from proteins \cite{nunes2018multibind}, conformational changes in enzymes \cite{zheng2018multikinase}, and base-pairing transitions in nucleic acids (see \cite{nikolova_2011_transient}, as well as \cite{perezdealbaortiz_2019_pmdbook} and \cite{perezdealba2021sequencedep}).
Here, we show how to extend our path-CV approach to search beyond the nearest local MFEP and, instead, explore multiple relevant reaction channels}.

Well-known path-CV schemes such as the string method in its different variants \cite{E2002:052301,E2005:6688,maragliano_2006_string} and the nudged elastic band method \cite{jonsson_1998_neb,Bohner2014_NEB} are able to locate an MFEP by propagating an initial guess path along the steepest descent on the free-energy landscape in the direction perpendicular to the path. In the string method, the path is expressed as a string of nodes, of which the first and last nodes are fixed at the stable-state minima. The other nodes can then be evolved by running a series of simulations restrained at each node position that accumulate  the gradient of the free energy for each path update. The initial path can simply be a linear interpolation between the stable states or a somewhat more informed initial guess. After the string of nodes has reached a local MFEP, the reaction free-energy profile can be computed as a function of the progress component along the converged path, $s$.

Branduardi and co-workers developed an approach that can be used to find additional reaction channels \cite{branduardi_2007_pathz}. They expressed a transition path as a series of ordered molecular structures that morph from the average reactant to the average product structure.
Next, they defined two components: the progress along the path, $s$, and the distance from the path, $z$. By performing a 2D metadynamics simulation in the ($s,z$)-space of an already optimized path, they could also search for other reaction channels.
The final 2D metadynamics free-energy surface would show the reference MFEP at $z$-values close to zero, and possible other paths at higher $z$-values. However, the higher $z$-distance from the optimized structures of the reference path does not provide structural information on the other mechanistic paths. Moreover, multiple paths at similar distances from the reference path cannot be distinguished. 

In this work, we present a method that finds multiple average transition paths (or local MFEPs) in a simultaneous manner.
\textcolor{black}{The method draws inspiration from multiple-walker \cite{raiteri_2006_mw} and replica-repulsion \cite{malevanets_2011_replicarepulsion} techniques, and builds on our path-metadynamics (PMD) approach, introduced in \cite{diazleines_2012_pmtd,perezdealbaortiz_2018_pmtdadv}.}
As the name suggests, PMD combines a path-CV with the metadynamics enhanced sampling technique \cite{laio_2002_mtd,Ensing2006_mtdreview,bussi_2020_mtdreview}.
By gradually building a repulsive bias potential along the path-CV, while simultaneously optimizing the path, PMD is able to escape stable states, find a (local) reaction pathway and obtain the free-energy profile along the converged path.
Instead of a sequence of molecular structures, the path-CV is expressed as a curve in the space of the CVs that each describe (part of) the transition.
The path optimization is particularly efficient because it is driven by the cumulative average density of the sampling and therefore does not need to sample the gradients of the free-energy landscape. 
If let unrestrained, PMD converges the path-CV to the average transition path; that is, to the maximum reactive flux.
However, if a stiff harmonic restraint, or ``tube'' potential, is set on the distance component perpendicular to the path, $z$, then PMD follows the local free-energy gradient as estimated from the narrow density fluctuations; converging to the MFEP.
\textcolor{black}{PMD has demonstrated subexponential scaling of the path and free-energy convergence time with respect to the growth of CV-space dimensionality \cite{perezdealbaortiz_2018_pmtdadv}.}
Recently, a multiple-walker implementation of PMD was realized, in which multiple replicas sample and optimize the same path-CV and free-energy profile in parallel \cite{perezdealbaortiz_2018_pmtdadv}.
Apart from a trivial speed-up in the free-energy calculation with the number of replicas, this also provides a more efficient path optimization due to the more evenly distributed sampling.

To find multiple reaction channels, we add an extra level of parallelism and concurrently spawn several PMD simulations (multiPMD), each of them with its own group of walkers that update their respective path-CV and free-energy estimation.
Naturally, when starting from the same initial guess path, all of these simulations would normally converge to the same transition mechanism.
However, this can be prevented by introducing repulsion between the path-CVs.
To do so, we introduce \textit{repellers}, a special type of walkers that does not participate in the free-energy calculation---i.e.\ they do not deposit Gaussian potentials, nor are biased by the potentials deposited by other walkers---but do take part in the path optimization and, additionally, exert a tunable repulsive potential on the repellers of the other paths.
While the repellers separate the path-CVs, the standard walkers perform a free-energy calculation along their respective paths.
\textcolor{black}{In the following, we show how repellers, and other types of special walkers, can successfully drive the search for separate reaction channels.
One main advantage of the multiPMD framework is that the walkers that are not repellers contribute explicit dynamics between the stable states, thus yielding physical paths within the simulation's level of theory.
This supposes an advantage over action-minimization strategies, which have been also used to locate multiple transition pathways \cite{fujisaki_2010_multiaction,lee_2017_multiaction,mandelli_2020_multirpmd}, but that may suffer from non-physical mechanisms arising from discretization errors \cite{lee_2020_multiactionprob}.
Another advantage of our proposed method is that the search for multiple paths can be conducted simultaneously. A recent multi-string version of the climbing string method was developed to locate saddles and pathways in a free-energy landscapes by combining static and dynamic strings \cite{shrivastav_2019_multiclimb}. Similarly, non-local perpendicular sampling, followed by node reordering via  a traveling-salesman scheme has been develop to find multiple transition pathways \cite{zhu2019taps}. However, both of this approaches require to first find and fix an initial path before searching for alternatives ones.
In addition, being implemented in PLUMED \cite{tribello_2014_plumed}, the multiPMD framework can be easily combined with other free-energy techniques besides metadynamics, and be used in combination with various other MD engines.}

\section{Methods}

\subsection{Multiple-path-metadynamics}
\begin{figure}[tbhp]
\centering
\includegraphics[width=.7\linewidth]{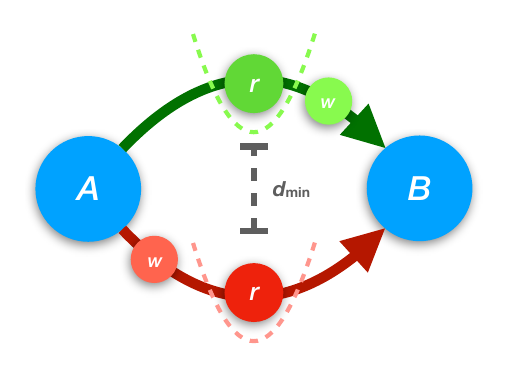}
\caption{Scheme representing the elements of the multiPMD method. Two stable states, $A$ and $B$, are represented with blue circles. Two adaptive paths connecting the two stable states are represented in green and red, respectively. At the midpoint of each path there is a repeller, $r$, fixed by a harmonic restraint represented with a dashed-line parabola. The two repellers exert a repulsive force on each other with a certain range, $d_\text{min}$, represented with a gray dashed line. Each path also has standard walkers, $w$, which perform a free-energy calculation.}
\label{chap4:fig1}
\end{figure}

In the simplest protocol, we initialize two parallel multiple-walker PMD runs, and we substitute one standard walker for a repeller in each of the two runs. The repellers are first steered to the midpoint of their respective path ($s$=0.5) by means of a moving harmonic restraint. At the same time, the repulsive potential is activated by a moving lower-half harmonic ``wall'' potential that biases the pairwise distance between the repellers in CV-space, starting at a value of $d=0$ and finishing at a desired minimum distance $d=d_\text{min}$, which forces the path-CVs to diverge (see Fig.~\ref{chap4:fig1}). The aim is to find two different reaction channels, which will typically be separated from each other by a divide in the free-energy landscape. The value of the minimum distance $d_\text{min}$ can therefore be tuned in such way that it is enough to push one repeller over such a dividing ridge, away from the other repeller. Once the two repellers are on opposite sides of a divide, they will spontaneously further separate as they descent into the different reaction channels. Note that it is not necessary to know the location of any free-energy divides beforehand; one can just monitor the switch in the repeller dynamics from driven by the repulsive potential to spontaneous separation, which marks the crossing of a divide. During and after the path-CV separation, the other walkers continuously overwrite and improve the metadynamics free-energy calculation \cite{ensing_2005_heal}, leading to converged profiles for each of the found transition pathways.

This minimalist protocol can easily be extended. If the number of reaction mechanisms is known beforehand, one can start with that many parallel multiple-walker PMD runs, each with a repeller that pushes away all other path-CVs, so that each reaction channel is found by a unique path-CV. Alternatively, the reaction channels can be found in a sequential manner, starting with one path-CV optimization, and continuing by adding one path-CV optimization at the time, thereby keeping the previously optimized path(s) fixed, 
\textcolor{black}{in a similar way to the recent multi-string version of the climbing string method \cite{shrivastav_2019_multiclimb}.}
The number of repellers per path-CV and their location along the path can be flexibly chosen, for instance to capture suspected bifurcation points. 
Other options include the distance metric between repellers, e.g., a simple Euclidean distance or a statistical variance, either on all or on some specific CVs; and the repulsive potential functional form, its parameters and its range of action, etc. Repellers can also push away from arbitrary points in CV-space, rather than other repellers.
Moreover, we can exploit other kinds of non-standard walkers.
For example, \textit{attractors}, which pinpoint known landmarks to capture in CV-space while keeping the rest of the path flexible. 
Several attractors and repellers can be used together. Finally, we emphasize that this protocol of parallel interacting path-CVs to find multiple reaction channels can straightforwardly be combined with various other enhanced sampling methods instead of metadynamics, for example: umbrella sampling \cite{torrie_1977_us}, constrained MD \cite{Carter1989:472}, or steered MD \cite{grubmuller_1996_smd,jarzynski_1997_smd}.

\subsection{PathMap}

When dealing with complex molecular transitions, one is typically not only interested in finding the multiple possible pathways, but also in understanding how the reaction channels merge or separate, and how the system switches between the mechanisms connecting different intermediate states.
With this in mind, once the transition pathways have been found via multiPMD, we propose a scheme to sample the free-energy surface between the optimized path-CVs in a two-dimensional landscape, or map, of the pathways.
This PathMap is spanned by an average progress parameter along the $N$ reaction channels, $\bar s$, and a metric of the closeness to each path, $\bar z$, in which we have labeled the paths with an integer $i$ and ordered them based on their relative proximity. Both parameters use a weighting, $w_i$, based on the proximity to each path:

\begin{align}
\bar s & = \sum\limits_{i}^{N} w_i s_i  \,\bigg/ \sum\limits_{i}^{N} w_i \label{chap4:eq1}\\
\bar z & = \sum\limits_{i}^{N} w_i i  \,\bigg/ \sum\limits_{i}^{N} w_i \\
w_i & =\frac{1- (z_i/r)^n}{1- (z_i/r)^m} \label{chap4:eq2}
\end{align}

\noindent
\textcolor{black}{where, $r$, $n$ and $m$ are the parameters of a switching function that assigns weights to the paths depending on proximity, as measured by their respective $z_i$ components.
The switching function parameters can be optimized according to the distances between the paths in CV-space.}

With the two new descriptors, we can use metadynamics to render a 2D free-energy landscape, $(\bar s,\bar z)$, which reveals the connectivity and branching of the pathways, as well as the barriers of switching mechanisms at any point between the states $A$ and $B$.
\textcolor{black}{Effectively, the PathMap compresses a $2N$-dimensional space, $(s_1,s_2,...,s_N,z_1,z_2,...,z_N)$, into a 2D one, $(\bar s, \bar z)$.
The idea to apply dimensionality reduction to analyze a landscape is not new, and is done in other algorithms such as diffusion-map \cite{Coifman7432,Singer16090,Ferguson13597}, sketch-map \cite{sketchmap2011,sketchmap2012}, isomap \cite{tenenbaum_2000_isomap,Das9885}, intrinsic map dynamics \cite{chiavazzo2017intrinsic}, and others \cite{Chen3235,Sims618,sultan_2017_tica,perezhernandez_2013_tica,tiwary_2016_sgoop}.}

\section{Results}
\subsection{Multiple transition channels of alanine dipeptide}
To illustrate the performance of multiPMD, we map the conformational transition channels of two well-known short peptides.
The first case study involves alanine dipeptide (Ace-Ala-Nme), a prototypical model system used for testing and benchmarking enhanced sampling methods.
Details on the simulation protocol, including system preparation and parameters for MD, metadynamics and the path-CVs, can be found in the Appendix~\ref{chap4:app}.
Runs are performed in GROMACS 2018.8 \cite{hess_2008_gromacs} patched with PLUMED 2.6 \cite{tribello_2014_plumed}.
The conformational transition of the dipeptide is described by two CVs, the dihedral angles $\phi$ and $\psi$.
Two path-CVs connect the stable states C7$_\text{eq}$ and C7$_\text{ax}$, starting with a linear interpolation.
Each path has three standard walkers and one repeller.
A tube potential is set on the direction perpendicular to the path, $z$, to keep the walkers close to their respective path and to keep the path updates small.
The tube's force constant is 70 kcal$\cdot$mol$^{-1}$$\cdot$rad$^{-2}$.
The repellers start each in one of the two stable states, and are steered to the midpoint of the path, i.e.\ $s=0.5$, with a force constant of 1500 kcal/mol per normalized path unit.
Such strong restraints are required, since we expect the repellers to sit close to the top of the free-energy barriers.
Simultaneously, the repulsion is activated by a lower harmonic wall that biases the squared distance between the repellers in CV-space, increasing from 0.0 to 0.5 rad$^2$, with a force constant of 70 kcal mol$^{-1}$ rad$^{-4}$.
To drive path separation successfully, the force constant of the repelling potential should be at least as strong as that of the tube potential.

\begin{figure}[tbhp]
\centering
\includegraphics[clip,width=0.6\linewidth]{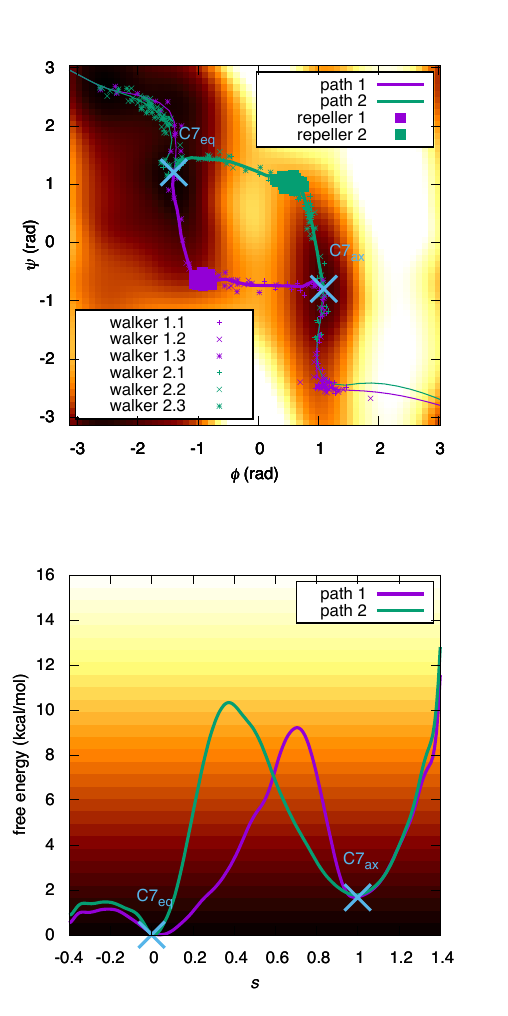}
\caption{Top: Two different C7$_{\text{eq}}\rightarrow$C7$_{\text{ax}}$ paths for alanine dipeptide captured by multiPMD are represented in purple and green. The sampling of repellers and standard walkers during the last 5 ps of simulation is shown with different markers of the same color as their respective paths. Bottom: Two free-energy profiles represented in the same color as their respective paths.}
\label{chap4:fig2}
\end{figure}

Fig.~\ref{chap4:fig2} shows the two well-separated paths after 10 ns of simulation.
At this stage, the repellers do not exert force on each other, as they are beyond the repulsion range (see Fig~\ref{chap4:figA1}).
Subsequently, the path-CVs relax to two local MFEPs.
The sampling performed by the metadynamics walkers is evenly distributed after a few ns of simulation (see Fig~\ref{chap4:figA1}), indicating good convergence.
The resulting free-energy profiles, shown in Fig.~\ref{chap4:fig2}, agree on the difference between the two stable states.
Path 1 shows a lower barrier, as expected for the dominant, global MFEP.

\textcolor{black}{The repellers and attractors can be used in a versatile and arbitrary manner to take further slices and probe interesting details of a high-dimensional free-energy landscape.
For example, the final positions of the two repellers in the $(\phi,\psi)$-plane can be used as initial and final points for a repeller-to-repeller path.
By placing attractors along different points of the repeller-to-repeller path, we can sample new C7$_{\text{eq}}\rightarrow$C7$_{\text{ax}}$ paths that ``switch'' from path 1 to path 2.
In Fig~\ref{chap4:figA2}, we show such switching paths, at one third, one half and two thirds of the distance from repeller 1 to repeller 2.
While the free-energy profiles along the switching paths are consistent with the $(\phi,\psi)$ free-energy surface, the repeller-to-repeller path is not guaranteed to provide a robust---or physically interpretable---switching parameter.
The PathMap approach offers a more general solution to sample between maps, which is illustrated for the polyproline case study hereafter.
More details about the switching-path approach can be found in the Appendix~\ref{chap4:app}.}

The alanine dipeptide system is also used to illustrate how to use an attractor for the treatment of a cyclic path (see Fig.~\ref{chap4:figA3}).
We set an C7$_{\text{eq}}\rightarrow$ C7$_{\text{eq}}$ path with an attractor at C7$_{\text{ax}}$, hence effectively tracking an C7$_{\text{eq}} \rightarrow $C7$_{\text{ax}} \rightarrow $C7$_{\text{eq}}$ path.
This approach does not require to make assumptions about the length of each section of the path in order to choose a node to fix at C7$_{\text{ax}}$.
It also exemplifies how an attractor can guide a path to a particular known intermediate state.
A more detailed account of this result is given in the Appendix~\ref{chap4:app}.
 
\subsection{Multiple transition channels of tetrameric polyproline}

The second peptide we study is a tetrameric polyproline (Ace-(Pro)$_4$-Nme).
See the Appendix~\ref{chap4:app} for a detailed description of the system preparation and parameters for MD, metadynamics and the path-CV.
We focus on the transition from the PPI right-handed helix to the PPII left-handed helix, which was also investigated in \cite{perezdealbaortiz_2018_pmtdadv}.
The conformation of the tetramer can be described by three CVs, the dihedral angles $\omega_1$, $\omega_2,$ and $\omega_3$ (numbered from the acetylated end), which have values of 0 rad at PPI and $\pm \pi$ at PPII.
The preferred transition mechanism is zipper like, starting with a counter-clockwise rotation (as defined in \cite{moradi_2009_abmd}) of $\omega_1$, followed by $\omega_2$ and $\omega_3$. 
However, a total of six different mechanisms are possible if all orders of counter-clockwise $\omega$-rotations are considered. 
We set to capture these mechanisms with six path-CVs.
Since we expect the transitions to have two intermediate states (after one and two $\omega$-rotations), we employ two repellers per path.
During the first 50 ps of the simulation, the repellers of each path are steered to $s=0.33$ and $s=0.66$ each, with force constants of 1500 kcal/mol per normalized path unit.
Simultaneously, each pair of repellers repels the average position of the five other pairs.
The repulsion is exerted by a lower harmonic wall that biases the squared distance between the average repeller-pair positions, increasing from 0.0 to 4.0 rad$^2$, with a force constant of 10 kcal mol$^{-1}$ rad$^{-4}$.
The repulsion is deactivated after 500 ps of simulation.
More details on the run parameters are available in the Appendix~\ref{chap4:app}.

Fig.~\ref{chap4:fig3} shows six distinct PPI$\rightarrow$PPII pathways identified by multiPMD after 2.5 ns.
Path 1 is the MFEP, as identified in \cite{moradi_2009_abmd} and \cite{perezdealbaortiz_2018_pmtdadv}.
The final paths are beyond the repulsive range, and relax to local MFEPs.
The time evolution of the repulsive force is shown in Fig.~\ref{chap4:figA4}.

\begin{figure}[tbhp]
\centering
\includegraphics[width=.8\linewidth]{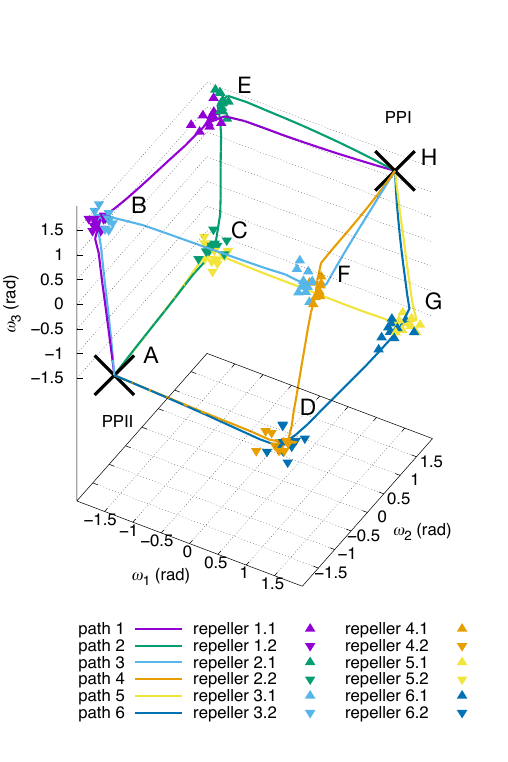}
\caption{Six different PPI$\rightarrow$PPII transition paths for tetrameric polyproline captured in 2.5 ns of simulation by multiPMD are represented in different colors. The sampling of repellers during the last 5 ps of simulation is shown with different markers of the same color as their respective paths. (Meta)stable states are labelled ``A'' to ``H'' to ease comparison with Fig.~\ref{chap4:fig4}.}
\label{chap4:fig3}
\end{figure}

\begin{figure}[tbhp]
\centering
\includegraphics[width=.8\linewidth]{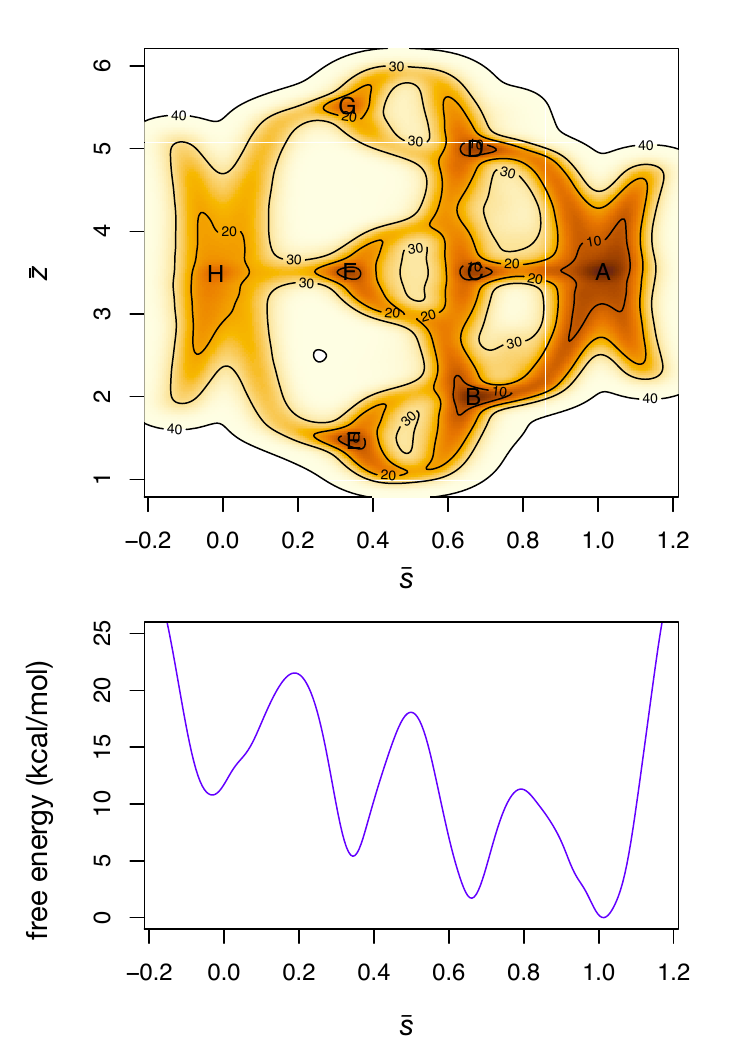}
\caption{Top: PathMap free-energy surface corresponding to the six PPI$\rightarrow$PPII paths for tetrameric polyproline shown in Fig.~\ref{chap4:fig3}. Bottom: PathMap free-energy projected onto $\bar s$. Both plots are generated with Metadynminer \cite{trapl_2020_metadynminer}.}
\label{chap4:fig4}
\end{figure}

To showcase how to sample the free-energy landscape between paths, we use the PathMap approach for the six located pathways of tetrameric polyproline.
We perform metadynamics on the $(\bar s,\bar z)$-plane, where $\bar s$ is the proximity-weighted progress along the six paths, and $\bar z$ is the closeness metric to each of the six paths.
The resulting PathMap free-energy surface is shown in Fig.~\ref{chap4:fig4}.
The minima are labelled ``A'' to ``H'', from the lowest to the highest free energy (See Table~\ref{chap4:tab1}).
Label ``H'' marks the PPI state, and label ``A'' marks the PPII state.
The intermediates with the lowest free energies, ``E'' and ``B'', correspond to the global MFEP, path 1. 
The features on the 2D PathMap match those of the paths in the full ($\omega_1,\omega_2,\omega_3$)-space.
Three channels connect the ``H'' state at $\bar s=0.0$, to the first intermediates, ``G'', ``F'' and ``E'', at $s\approx0.33$.
This structure reflects the overlap of the paths---1 with 2, 3 with 4, and 5 with 6---in Fig.~\ref{chap4:fig3}.
As the transition advances, each of the three channels bifurcates, leading to the six channels that are distinguishable at $\bar s\approx0.5$, and also in Fig.~\ref{chap4:fig3}. 
Forward in the transition, at $\bar s \approx0.66$, the six channels merge again into three, corresponding to the overlap of the paths 1 with 3, 2 with 5 and 4 with 6 in Fig.~\ref{chap4:fig3}.
Finally, the three channels connect the intermediates ``D'', ``C'' and ``B'' to state ``A'' at $\bar s=1.0$. 
It is also relevant to notice that the free energy of the intermediates is lower for $\bar z$ close to path 1, and higher as $\bar z$ approaches path 6.
This PathMap projection is a consequence of our careful ordering of the paths, but future versions of the method can contemplate an automated path indexation.
The projection of the free-energy along the $\bar s$ component shows the three barriers corresponding to the three $\omega$-rotations, similar to the result in \cite{perezdealbaortiz_2018_pmtdadv}.
 
\section{Conclusions}
To conclude, here we have shown how path-CVs guided by repulsive walkers can simultaneously find different molecular transition pathways.
The repellers can be flexibly tuned in number, biasing potential and range, or even act as attractors.
Once these special walkers force the multiple path-CVs to diverge, the standard walkers can converge a free-energy profile for each mechanism with high-efficiency, independent of the dimensionality of the CV-space.
After the pathways have been isolated, the free energy of a foliation between two, or more, pathways can be calculated in a 2D metadynamics run, or PathMap, which can be a useful measurement in diffusive transitions.

Current large-scale computing facilities offer excellent platforms for parallel replica simulations, which multiPMD can exploit.
By implementing the repulsion via special walkers, i.e., repellers, we ensure that they obey the potential-energy surface of the system, apart from the biasing force.
Another advantage of assigning the repulsion, or attraction, to specific replicas is that the remaining walkers continue to perform a standard free-energy calculation, i.e., metadynamics, along the evolving path.
The dynamics of the standard walkers, which cross and recross from $A$ to $B$ numerous times, yields physical transition paths within the level of theory of the simulation, free from time discretization errors.
Each of the repelled paths is optimized to a local MFEP, which can be related to a specific contribution to the total rate constant of the transition, given by their free-energy barrier \cite{zaman_2003_rates}.

\textcolor{black}{In \cite{perezdealba2021sequencedep}, multiPMD is employed for an investigation of base-pairing transition pathways in DNA.}
The current framework also sets the stage for future developments.
For example, in the current implementation, the location of the repellers along the paths has to be chosen a priori. 
A climbing algorithm could potentially guide the repellers to candidate transition states \cite{shrivastav_2019_multiclimb}.
Additionally, instead of setting the number of paths and repellers at the beginning of the simulation, new paths could be initialized depending on specific convergence criteria. 
Another prerequisite of multiPMD is the definition of the CV-space itself. 
CV discovery techniques based on machine learning have been successfully combined with path-CVs to describe phase transformations \cite{rogal_2019_nnpath}.
Hence, multiPMD is not only a ready-to-run flexible method that can deliver multiple physical molecular transition pathways, but also a scheme with promising directions for the near future. 

All the data and PLUMED input files required to reproduce the results reported in this paper are available on PLUMED-NEST (\url{www.plumed-nest.org}), the public repository of the PLUMED consortium \cite{bonomi_2019_plumednest}, as plumID:21.049.

\FloatBarrier

\section{Appendix}
\label{chap4:app}

\renewcommand{\thefigure}{A\arabic{figure}}
\renewcommand{\thetable}{A\arabic{table}}

\subsection{Multiple-path-metadynamics of alanine dipeptide}
The peptide C7$_\text{eq}$ configuration is generated with AmberTools17 \cite{ambertools_2017}.
The C7$_\text{ax}$ state is obtained in a short path-steered MD run \cite{grubmuller_1996_smd}, with similar parameters as the ones used in production runs.
Interatomic interactions are modeled by the  AMBER99SB-ILDN force field \cite{lindorff_2010_amberff99sbildn}.
We run 50 ps of equilibration at each stable state, C7$_\text{eq}$ and C7$_\text{ax}$, using GROMACS 2018.8 \cite{hess_2008_gromacs}.
The time step is 2 fs.
The LINCS algorithm \cite{hess1997lincs} is used to constrain bonds involving hydrogen atoms.
We use the canonical sampling through velocity rescaling (CSVR) thermostat at 300 K with a time constant of 0.1 ps \cite{bussi_2007_csvr}.
The same settings are kept for the production runs.

We use PLUMED 2.6 \cite{tribello_2014_plumed} with the adaptive path-CV code available at \url{http://www.acmm.nl/ensing/software/PathCV.cpp} to perform multiPMD runs.
The CVs are the dihedral angles $\phi$ and $\psi$.
The stable states are defined: C7$_\text{eq}$ (state $A$) at ($\phi=-1.4 \text{ rad},\psi=1.2 \text{ rad}$), and C7$_\text{ax}$ (state $B$) at $(\phi=1.1 \text{ rad},\psi=-0.8 \text{ rad})$.
We set harmonic walls with force constants of 1000 kcal$\cdot$mol$^{-1}$$\cdot$rad$^{-2}$ to keep sampling in the interval $\phi=[-2.7,2.7]$ and $\phi=[-2.7,2.7]$.
The two paths consist each of 60 nodes in total: 20 nodes between the stable states $A$ and $B$ to cover the transition, 20 extra nodes before, and 20 extra nodes after.
The extra nodes help in sampling slightly past the bottom of the free-energy valleys.
The paths are initialized as a linear interpolation between $A$ and $B$ and are updated every 0.5 ps according to the algorithm described in \cite{diazleines_2012_pmtd}. 
The half-life parameter, which controls path flexibility, is set to 2 ps. 
This is the time after which previous samples  weight only half of their original value for the current path update.
A tube potential is set on the direction perpendicular to the path, $z$, to keep the walkers close to their path and the path updates small.
The tube's force constant is 70 kcal$\cdot$mol$^{-1}$$\cdot$rad$^{-2}$.
Similarly, we put walls on the direction of the path, $s$, at $-0.4$ and 1.4 with a force constant of 50 kcal/mol per normalized path unit to keep the sampling within the relevant $s\approx[0,1]$ range.
For more details on how path parameters are chosen, consult \cite{perezdealbaortiz_2019_pmdbook}.

\begin{figure}
\centering
\includegraphics[width=0.6\columnwidth]{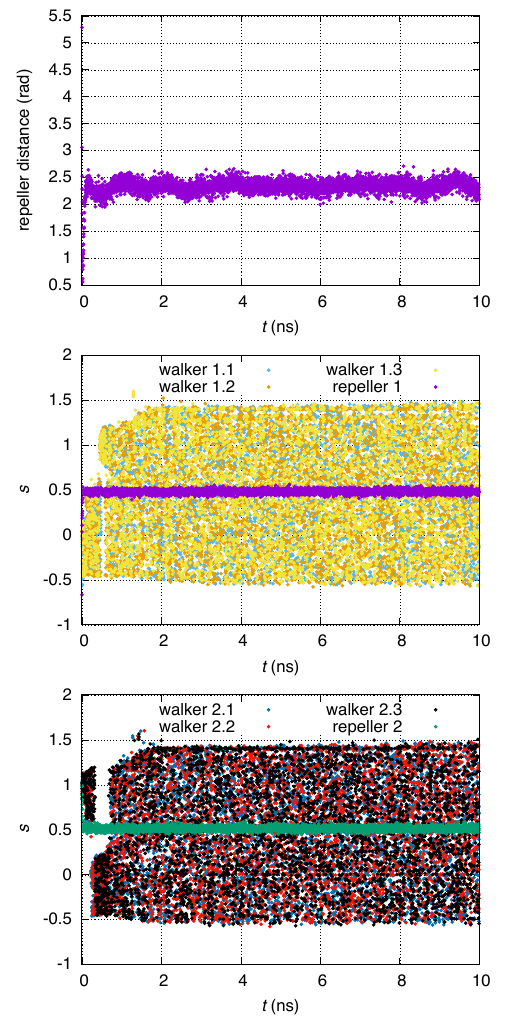}
\caption{\label{chap4:figA1} Top: time series of distance between repellers of the two C7$_\text{eq}$ $\rightarrow$ C7$_\text{ax}$ paths for alanine dipeptide. Middle and Bottom: Time series of the sampling along each of the two C7$_\text{eq}$ $\rightarrow$ C7$_\text{ax}$ paths for alanine dipeptide by the standard walkers and repellers.}
\end{figure}

The repulsion parameters are provided in the main text.
The location in CV-space of each walker is communicated to the other walkers via weighted replica averages in PLUMED.
Fig.~\ref{chap4:figA1} depicts the distance between the repellers.
After a few ps, the repellers are beyond each other's range and do not exert force.
This indicates that, once separated, the paths evolve obeying the undisturbed free-energy surface of the system.

The standard walkers that perform metadynamics (three on each path) deposit Gaussian potentials with a height of 0.02 kcal/mol and a width of 0.05 normalized path units every 0.5 ps.
The Gaussians are stored on a grid for efficiency, and only exert force within the $s=[-0.4,1.4]$ interval, to avoid artifacts with the harmonic walls limiting that same range.
The sampling performed by the metadynamics walkers is shown in Fig.~\ref{chap4:figA1}.
The evenly distributed sampling indicates good convergence in the free-energy calculation.
A free-energy profile is obtained by averaging the metadynamics estimations, every 50 ps, from 1 to 10 ns.

\subsection{Switching-path-metadynamics of alanine dipeptide}

\begin{figure}
\centering
\includegraphics[width=0.6\columnwidth]{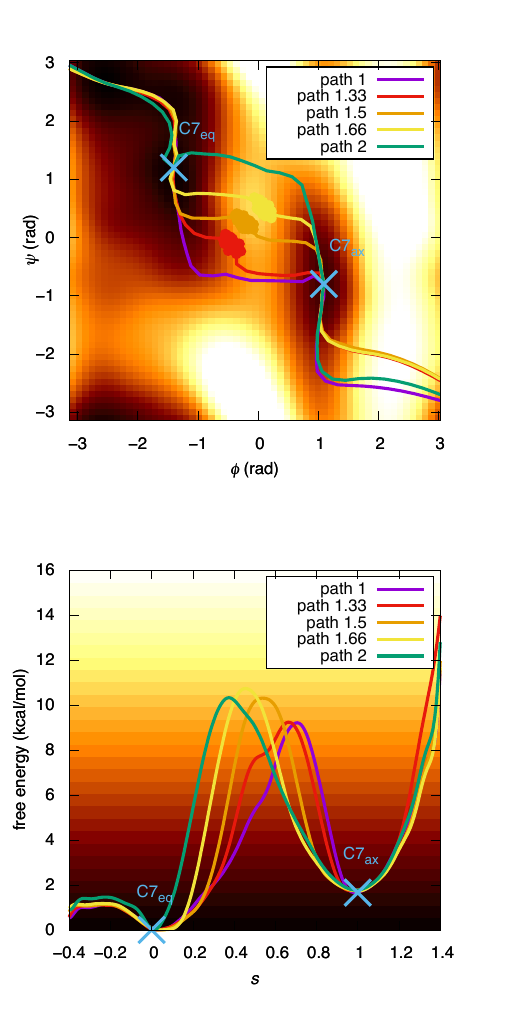}
\caption{\label{chap4:figA2} Top: Three different C7$_{\text{eq}}\rightarrow$C7$_{\text{ax}}$ paths for alanine dipeptide progressively switching from path 1 to path 2, as shown in Fig.~\ref{chap4:fig2}. The intermediate states of each of the three paths are defined by attractors sitting between the final positions of repeller 1 and repeller 2, as shown in Fig.~\ref{chap4:fig2}. The three paths are labeled $1 + s'$, where $s'$ is the progress in their position from repeller 1 to repeller 2. The sampling of attractors during the last 5 ps of simulation is shown with markers of the same color as their respective paths. Bottom: Free-energy profiles represented in the same color as their respective paths.}
\end{figure}

\textcolor{black}{We extract the final positions of the two repellers from the alanine dipeptide multiPMD run.
The position of repeller 1 is ($\phi=-1.0 \text{ rad},\psi=-0.6 \text{ rad}$), and of repeller 2 is ($\phi=0.7 \text{ rad},\psi=1.0 \text{ rad}$). 
We define a fixed path from repeller 1 to repeller 2 with a linear interpolation of 10 nodes.
We refer the progress component along this repeller-to-repeller path as $s'$.
We perform a PMD run from C7$_\text{eq}$ to C7$_\text{ax}$ with three standard walkers and one attractor.
The standard walkers perform metadynamics with the same parameters as those of the multiPMD run.
The attractor is restrained to $s=0.5$ along the C7$_\text{eq}$-to-C7$_\text{ax}$ path, and to $s'=0.5$ along the repeller-to-repeller path. 
Both force constants are of 1500 kcal mol$^{-1}$ units per squared normalized path unit.
The restraints are stiff because we expect the attractor to sit on top of the free-energy barriers.
We then repeat two similar runs, but with the attractor restrained at $s'=0.33$ and $s'=0.66$ of the repeller-to-repeller path.
The resulting pathways, gradually switching from path 1 to path 2, are shown in Fig.~\ref{chap4:figA2}.
We obtain free-energy profiles in the same manner as for the multiPMD run.}

\subsection{Cyclic-path-metadynamics of alanine dipeptide}

The CVs are $\cos(\phi), \sin(\phi), \cos(\psi), \sin(\psi)$.
We initialize a cyclic path with 39 nodes from C7$_\text{eq}$ to C7$_\text{eq}$ ($\phi=-1.4 \text{ rad},\psi=1.2 \text{ rad}$).
The path initially describes a concerted rotation of both dihedral angles.
The path is updated every 0.5 ps, with a half-life of 2 ps.
The tube potential has a force constant of 50 kcal$\cdot$mol$^{-1}$$\cdot$rad$^{-2}$.
The path progress parameter, $s$, is scaled, shifted and handled periodically from $s=[-1,1]$, in such way that $s=\pm 1$ marks the C7$_\text{eq}$ state. 
We expect the C7$_\text{ax}$ state to be at $s \approx 0$, but the exact value cannot be known a priori, since it depends on the length of each section of the cyclic path.
An attractor is initialized at C7$_\text{ax}$ $(\phi=1.1 \text{ rad},\psi=-0.8 \text{ rad})$, with force constants of 50 kcal$\cdot$mol$^{-1}$$\cdot$rad$^{-2}$ on each CV.
Seven walkers perform metadynamics with Gaussian height 0.08 kcal/mol and width 0.05 path units, deposited every 0.5 ps.
A free-energy profile is obtained by averaging the metadynamics estimations, every 50 ps, from 1 to 10 ns.

\begin{figure}
\centering
\includegraphics[width=0.6\columnwidth]{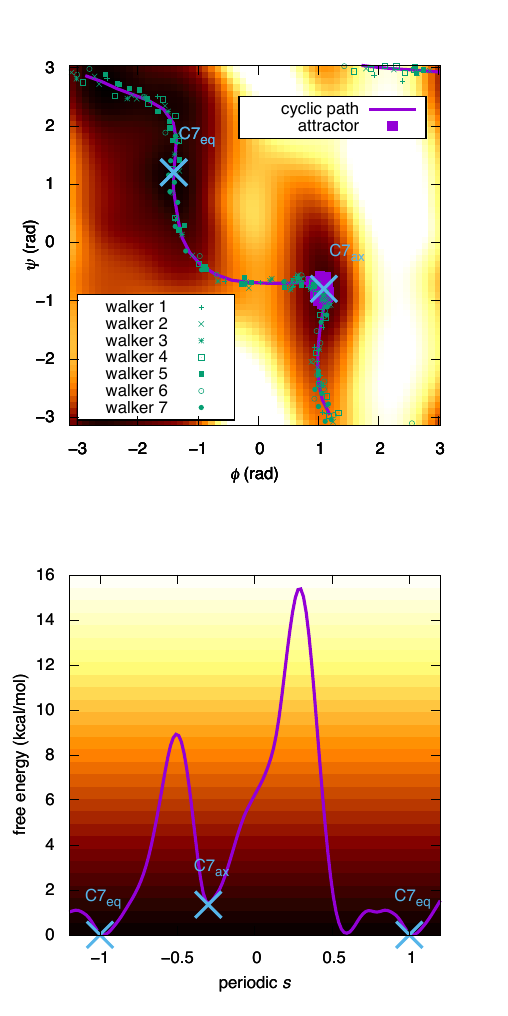}
\caption{\label{chap4:figA3} Top: A C7$_{\text{eq}}\rightarrow$C7$_{\text{ax}}\rightarrow$C7$_{\text{eq}}$ cyclic path  for alanine dipeptide captured in 10 ns by cyclicPMD of simulation. The sampling of attractor and standard walkers is shown during the last 5 ps of simulation. Bottom: Free-energy profile along the cyclic path.}
\end{figure}

Fig.~\ref{chap4:figA3} shows the cyclic path projected on the $(\phi,\psi)$-plane.
The first part of the cyclic path is the same as path A depicted in Fig.~\ref{chap4:fig2}.
Its free-energy profile is consistent with the previous result.
The second part corresponds to a third path that completes a periodic crossing in the depicted ($\phi,\psi$)-plane.
The attractor ensures that the path passes through the C7$_\text{ax}$ state, while allowing for flexibility on the length of each path section (i.e., before and after the C7$_\text{ax}$ state).
In fact, in the free-energy profile, we observe that the first part of the path, corresponding to $s=-1$ to $s=-0.3$, is shorter that the second part, from $s=-0.3$ to $s=+1$.

\subsection{Multiple-path-metadynamics of tetrameric polyproline}

Similar to alanine dipeptide, the peptide PPII configuration is generated with AmberTools17 \cite{ambertools_2017}.
The PPI state is obtained via a short path-steered MD run \cite{grubmuller_1996_smd}, with similar parameters as the ones used in production runs.
We use the AMBER force field ff99SB \cite{hornak_2006_amberff99sb} to model interatomic interactions.
50 ps of equilibration are performed at each stable state, PPI and PPII, in GROMACS 2018.8 \cite{hess_2008_gromacs}.
The time step size is 1 fs and a canonical sampling through velocity rescaling (CSVR) thermostat \cite{bussi_2007_csvr} is set at 300 K with a time constant of 0.1 ps.
The same parameters are used for the production runs.


Same as for alanine dipeptide, we use PLUMED 2.6 \cite{tribello_2014_plumed} and the adaptive path-CV code available at \url{http://www.acmm.nl/ensing/software/PathCV.cpp}.
The stable states are: PPI ($A$) at ($\omega_1=0 \text{ rad},\omega_2=0 \text{ rad},\omega_3=0 \text{ rad})$ and PPII ($B$) at ($\omega_1=\pm\pi \text{ rad},\omega_2=\pm\pi \text{ rad},\omega_3=\pm\pi \text{ rad})$.
The CVs that are included in the path are $\omega'_i = \omega_i + \pi/2$, for $i=1,2,3$, to ease periodic treatment as done in \cite{perezdealbaortiz_2018_pmtdadv}.
We set harmonic walls with force constants of 1000 kcal$\cdot$mol$^{-1}$$\cdot$rad$^{-2}$ to keep sampling in the interval $\omega'_i=[-2.9,2.9]$.
The six paths have each 30 nodes in total, all between $A$ and $B$ to cover the transition.
No extra nodes are added.
The paths are initialized as a linear interpolation between $A$ and $B$ and are updated every 0.5 ps according to the algorithm described in \cite{diazleines_2012_pmtd}. 
The half-life parameter, which controls path flexibility, is set to 2.5 ps. 
The force constant of the tube potential is 10 kcal$\cdot$mol$^{-1}$$\cdot$rad$^{-2}$.

\begin{figure}
\centering
\includegraphics[width=0.6\columnwidth]{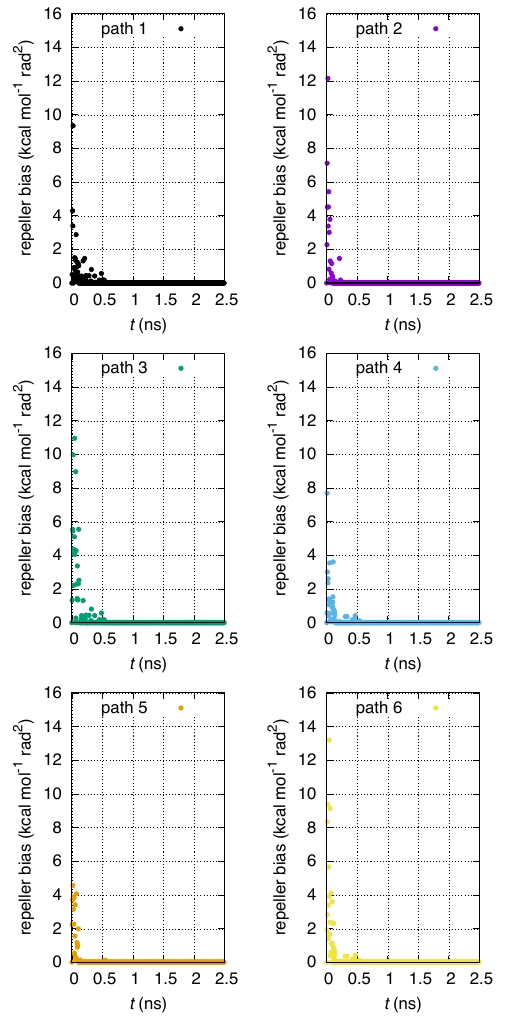}
\caption{\label{chap4:figA4} Time series of the instantaneous value of the repulsive potential for the repellers of each of the six PPI $\rightarrow$ PPII paths for tetrameric polyproline.}
\end{figure}

The repulsion details are provided in the main text.
In Fig.~\ref{chap4:figA4} we show the time evolution of the repulsive bias exerted on the repellers.
At 0.55 ns, the repulsion is turned off.
By that time, the paths have been successfully separated, and relax to local MFEP beyond each other's repulsive range. 

\subsection{PathMap of tetrameric polyproline}
\textcolor{black}{We start by downsampling the six paths obtained in the multiPMD run of tetrameric polyproline.
We simplify the paths to just four nodes, which mark the PPI state, their two respective intermediate states, and the PPII state.
In order to calculate $\bar s$ and $\bar z$, we use Equations~\ref{chap4:eq1} to~\ref{chap4:eq2}, with $N=6$, $r=1$, $n=8$ and $m=16$. 
The parameter of the switching function ($r,n,m$) are carefully chosen by trial and error on a simplified version of the system---i.e.\ an analytical potential energy surface---to save computational time.
The paths are carefully ordered to achieve the overlap patterns described in the main text.}

\textcolor{black}{We perform metadynamics with 12 walkers on the $(\bar s, \bar z)$-plane with Gaussians of width $(0.05,0.1)$ and height 0.05 kcal/mol, deposited every 1 ps.
Wall potentials with force constants of 100 kcal$\cdot$mol$^{-1}$$\cdot$rad$^{-2}$ are used to keep the sampling in the interval $\omega'_i=[-2.6,2.6]$.
We present the free-energy estimation after 4 ns.
The metadynamics Gaussians are processed using Metadynminer \cite{trapl_2020_metadynminer}.
In Table~\ref{chap4:tab1}, we show the identified minima in the PathMap free-energy landscape.}

\FloatBarrier

\begin{table}[h]
\centering
\caption{Free-energy minima in the PathMap from Fig.~\ref{chap4:fig4}.}
\resizebox{.7\linewidth}{!}{%
\begin{tabular}{l c c c}
minima  & $\bar s$ & $\bar z$ & free energy (kcal/mol)  \\
\hline
A (PPII)    & 1.01 & 3.51  & 0.00\\
B    & 0.67 & 2.00  &    1.56 \\
C    & 0.67 & 3.51  &   5.75 \\
D    & 0.67 & 5.00  &   5.79 \\
E    & 0.35 & 1.46  &   6.29 \\
F    & 0.34 & 3.51  &   7.13 \\
G    & 0.33 & 5.51  &  11.31 \\
H (PPI)   & -0.02 & 3.49 &   12.65 \\
\end{tabular}
\label{chap4:tab1}
}
\end{table}

\section{Acknowledgements}

Simulations are performed on the carbon cluster at the University of Amsterdam;  A.P.A.O. received funding from the Mexican National Council for Science and Technology (CONACYT).

\newpage

\bibliographystyle{plain}
\bibliography{main.bib}

\end{document}